\documentclass[aps,reprint,showpacs,preprintnumbers,amsmath,amssymb,prb,citeautoscript]{revtex4-1} 

\usepackage{bm}
\usepackage{color}
\usepackage{dcolumn}
\usepackage{graphicx}
\usepackage[export]{adjustbox}
\graphicspath{{figs/}{./}}
\newcommand\FigureFile[1] {#1.eps}
\DeclareGraphicsExtensions{pdf}

\newcommand\eq[1]                              
{
\begin{align*}
#1
\end{align*}
}

\newcommand\eql[2] 
{
\begin{equation}\label{#1}
\begin{split}
#2
\end{split}
\end{equation}
}

\newcommand\eqsl[1]                            
{
\begin{align}
#1
\end{align}
}

\newcommand\eqssl[2]                      
{
\begin{subequations}\label{#1}
\begin{align}
#2
\end{align}
\end{subequations}
}

\newcommand\Fig[1]     {Fig.~\ref{#1}}

\newcommand\Ref[1]     {Ref.~\onlinecite{#1}}

\newcommand\compPackage[1] {{\footnotesize{#1}}}

\newcommand\NWCHEM     {\compPackage{NWCHEM}}

\newcommand\Eh[1][]    {\ensuremath{E_\mathrm{h}#1}}
\newcommand\mEh[1][]   {\textrm{m}\ensuremath{E_\mathrm{h}#1}}

\newcommand\Order[1]   {\mathcal{O}\left(#1\right)}

\newcommand\Crtwo      {\textrm{Cr$_{\textrm{2}}$}}
\newcommand\Motwo      {\textrm{Mo$_{\textrm{2}}$}}

\definecolor{xmgrace-green4}{rgb}{0.0,0.55,0.0}
\definecolor{Green}{rgb}{0.2,0.96,0.2}
\definecolor{Remarks}{rgb}{1,0.3,0.3}
\definecolor{Extra}{rgb}{0.2,0.2,1}
\definecolor{Blue}{rgb}{0.2,0.3,1}
\definecolor{Black}{rgb}{0,0,0}

\newcommand\COMMENTED[1] {}

\begin{document}

\title{Auxiliary-field quantum Monte Carlo calculations of the molybdenum dimer}

\author{Wirawan Purwanto}
\altaffiliation[Present affiliation: ]
{Old Dominion University, Norfolk, Virginia 23529, USA}
\email{wirawan0@gmail.com}
\affiliation{Department of Physics, College of William and Mary,
Williamsburg, Virginia 23187-8795, USA}

\author{Shiwei Zhang}
\affiliation{Department of Physics, College of William and Mary,
Williamsburg, Virginia 23187-8795, USA}

\author{Henry Krakauer}
\affiliation{Department of Physics, College of William and Mary,
Williamsburg, Virginia 23187-8795, USA}

\date{\today}

\begin{abstract}
Chemical accuracy is difficult to achieve for systems with transition metal atoms. Third row transition metal atoms are particularly
challenging
due to strong electron-electron correlation in localized $d$-orbitals.
The {\Crtwo} molecule is an outstanding
example, which we previously treated
with highly accurate
auxiliary-field quantum Monte Carlo (AFQMC) calculations
[Purwanto {\em et al., J. Chem. Phys.} {142}, 064302 (2015)].
Somewhat surprisingly, computational description of
the isoelectronic {\Motwo} dimer
has also, to date, been scattered and less than satisfactory.
We present high-level
theoretical benchmarks of the {\Motwo} singlet ground
state ($X ^1\Sigma_g^+$) and first triplet excited state ($a ^3\Sigma_u^+$),
using the phaseless AFQMC calculations.
Extrapolation to the complete basis set limit is performed.
Excellent agreement with experimental spectroscopic constants is obtained.
We also present a comparison of
the correlation effects in {\Crtwo} and  {\Motwo}.

\end{abstract}


\maketitle

\section{Introduction}

Transition metal (TM) dimers are of special interest theoretically and computationally.
They fully exhibit the complexity of more complex TM materials (such as the
formation of high-order bonds), but
their relatively small sizes make them amenable to systematic and rigorous
theoretical studies.
Group VIB dimers are especially interesting,
because the atom fragments are in the high-spin state ($^7$S), and they
form a closed shell ($^1\Sigma$) configuration in the molecular ground state.
This results in
a many-body spectrum with many nearly degenerate states,
with
strong electronic correlation effects.
The
{\Motwo} molecule is similar to {\Crtwo} in that both are highly
multiconfigurational in nature and require accurate treatment of both
static and dynamic electron correlation.
Because the 4d orbitals are not as localized as 3d, the severity of
electronic correlation is significantly
reduced for {\Motwo}.
For example, a complete active space self-consistent field (CASSCF) treatment
in {\Motwo} recovers about $30\%$
of the experimental binding energy,
while with the same wave function, {\Crtwo} is not even bound.
Nevertheless, the best quantum chemistry calculations for {\Motwo}
give
widely varying
predictions \cite{Balasubramanian2002,Angeli2007,Borin2008}.

In this work, we present accurate theoretical calculations of
{\Motwo} potential energy curves (PECs) near the equilibrium geometry.
We consider both the singlet ground state ($X\,^1\Sigma_g^+$)
and triplet first excited state ($a\,^3\Sigma_u^+$),
and calculate their spectroscopic properties.
We employ the phaseless auxiliary-field quantum Monte Carlo (AFQMC)
method
\cite{Zhang2003,AlSaidi2006b,Suewattana2007,Zhang2013}
in our calculations.
Select benchmark calculations were also performed with
exact free-projection (FP) AFQMC \cite{Zhang2013,Purwanto2009_Si,Shi2013}
to help establish the accuracy of our calculations.
A high-quality quadruple zeta (QZ) basis set is used for the majority of our calculations.
Extrapolation to the complete basis set (CBS) limit is performed in combination with
the triple zeta (TZ) basis results, following standard approaches.
As shown below, the AFQMC results for the spectroscopic constants are in
excellent agreement with experiment.
We compare our results with those from other quantum chemistry approaches.
The similarities and differences in electron-electron correlation between
{\Motwo} and {\Crtwo} are also examined.

The remainder of the paper proceeds as follows. The methodology is discussed in Section \ref{sec:Method}.
Results of our {\Motwo}  calculations
are presented in
Section \ref{sec:Results}.
Section \ref{sec:Discussion} presents comparisons with previous many-body quantum chemistry results and
an analysis of the relative
sizes of the correlation energy contributions in {\Motwo} and {\Crtwo},
as well as the effect of the trial wave function in the AFQMC calculations.
We summarize our results in Section \ref{sec:Summary}.

\section{Methodology}
\label{sec:Method}

The AFQMC method projects the many-body ground state wave function from a
given trial wave function.\cite{Zhang1997_CPMC,Purwanto2004,Purwanto2005}
It is implemented as random walks of Slater determinants, with orbitals expressed in
a chosen single-particle basis.
A phaseless approximation\cite{Zhang2003} has been introduced to control
the phase problem introduced by complex auxiliary-field sampling,
resulting in a practical computational method
that scales modestly with the system size
[e.g. $\Order{N^3}$ or $\Order{N^4}$].
Its high accuracy has been demonstrated in many molecular and solid systems
\cite{AlSaidi2006b, Purwanto2009_C2, Purwanto2009_Si,
Virgus2014, Purwanto2015_Cr2, Ma2015}
as well as model
electronic systems.
\cite{Shi2014,Zhang2013}

As an orbitally based wave function method, the AFQMC theoretical framework has close relations to many-body quantum chemistry methods. When expressed in a one-particle gaussian type orbital (GTO) basis, both approaches use exactly the same Hamiltonian. Thus, many efficient techniques developed for correlated quantum chemistry methods can be directly imported. This was done, for example, by using Cholesky decomposition to remove a bottleneck in the handling of two-body interaction matrix elements for large basis sets \cite{Purwanto2011} and introducing frozen-core type of approaches to seamlessly embed the highly correlated
AFQMC treatment within a larger mean-field environment \cite{Virgus2014,Ma2015}.

Since static correlation is important in {\Motwo},
it is desirable that the trial wave function account for this effect well.
We perform
CASSCF(12e,12o)
calculations, which correlate 12 electrons in 12 active orbitals.
The resulting multideterminant expansion is then truncated to retain about $93-95\%$
of the total weight, yielding a multi-determinant wave function which is used without
 further optimization as our trial wave function.
Typically this gives a trial wave function with $90-240$ determinants for the
ground state, and $30-290$ for the excited state.
Since the number of determinants grows rapidly with bond
stretching, we have to use a shorter cutoff
($86-92\%$) at larger bond lengths ($\ge 2.2$\,\AA) in order to keep the
number of determinants in a range that is easy to handle with the current state of our code.

We perform our AFQMC calculations using the all-electron
atomic natural orbital relativistic correlation consistent (ANO RCC)
GTO basis.\cite{Roos2005}
Scalar relativity
is treated with the Douglas-Kroll-Hess
Hamiltonian.
Spin-orbit effects on the dissociation energy were assessed with density functional
theory calculations,
using {\NWCHEM},
with a cc-pwCVTZ-PP basis and Dolg's energy-consistent spin-orbit relativistic effective core potential (ECP).
There was essentially no effect on the dissociation energy (with the spin-orbit generalized gradient approximation PBE functional), so
this was not pursued further.
The Ar+3d\textsuperscript{10} atom-like core orbitals are frozen at the
mean-field (CASSCF) level.
The calculated results were obtained using basis sets up to the realistic QZ
(8s7p5d3f2g1h contraction)
basis \cite{Borin2008}, denoted ANO-QZ hereafter.
As shown below,
this basis provides excellent results for many quantities.
The exception is the binding energy, which needs to be extrapolated to the
CBS
limit in order to make reliable comparisons with experiment.

\section{Results}
\label{sec:Results}

We first present  {\Motwo}
AFQMC results obtained with the ANO-QZ basis, before discussing
the CBS-extrapolated results.
Previously reported many-body calculations\cite{Balasubramanian2002,Angeli2007,Borin2008} largely used
multireference perturbative methods.
The  ANO-QZ CASPT2  (complete active space second-order perturbation theory)
results of Borin {\em et al.}\cite{Borin2008}  are among the most accurate.
The  exact and approximate AFQMC results, using the same basis, will provide a useful benchmark
to these results.

\begin{figure}[thbp]
\includegraphics[scale=0.35]{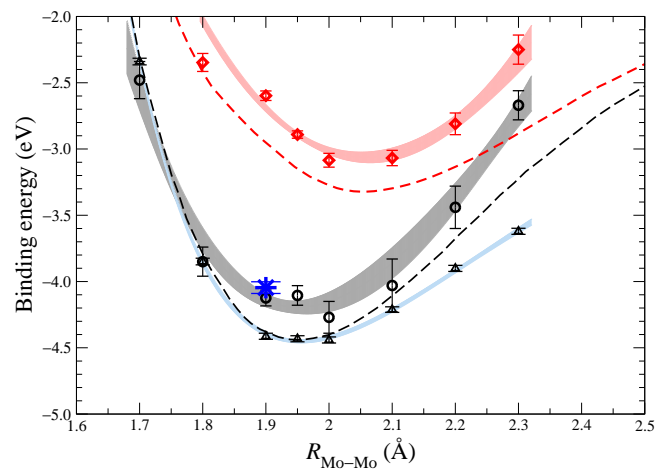}

\caption{\label{fig:Mo2-PEC}
AFQMC and CASPT2 {\Motwo} PECs, both using the ANO-QZ basis, for the ground state (dark colors)
and the lowest-energy triplet excited state (red).
Shaded curves are Morse fits to the AFQMC results.
CASPT2 PECs from \Ref{Borin2008} are shown as dashed lines.
Ground-state AFQMC: AFQMC/UHF results are shown as triangle symbols with error bars and light blue shading;
AFQMC/CASSCF are shown as black circles and grey shading. Only AFQMC/CASSCF results are shown for
the excited state (red diamonds and red shading). Exact \mbox{FP-AFQMC} result  for the ground state at \mbox{$R($Mo--Mo$) = 1.9$\,\AA}  is shown
by the blue star symbol.
}
\end{figure}

Figure~\ref{fig:Mo2-PEC} shows the {\Motwo} AFQMC
binding-energy curves of the
ground (singlet $X$) and excited (triplet $a$) states in the ANO-QZ basis.
(The binding energy is defined as the difference between
the molecular total energy and that of the two isolated atoms.
It is shown as a function of $R_\textrm{Mo--Mo}$, the distance between the two
{\Motwo} nuclei.)
AFQMC calculations of the Mo atoms were done using the ROHF trial wave function.
For the ground state of the molecule, phaseless AFQMC calculations are performed using both UHF \mbox{(AFQMC/UHF)} and
truncated CASSCF \mbox{(AFQMC/CASSCF)}
trial wave functions.
The computed binding energy curves are shown; also shown is the exact free projection AFQMC \mbox{(FP-AFQMC)} binding energy for a
geometry near equilibrium.
For the excited state, only AFQMC/CASSCF results are shown.
Morse curves are fitted to these results and are shown
as color bands whose width represents the combined stochastic and fitting uncertainties.
Near the experimental equilibrium geometry,
the ground state appears to exhibit stronger static correlation,
as
evidenced by the larger statistical error bar for the same amount of AFQMC computation.
This is also consistent with the observation that the Hartree-Fock energy for the singlet state is higher
than the triplet and that
a larger number of determinants are in the trial wave function for the ground state even though the
same cutoff is applied when truncating both the ground- and excited-state CASSCF wave functions (see Sec.~\ref{sec:Method}).

While AFQMC/UHF overestimates the binding energy by $\sim 0.25$\,eV compared to the exact FP-AFQMC,
AFQMC/CASSCF shows excellent agreement.
This establishes the high accuracy of the truncated CASSCF trial wave function for {\Motwo}.
Similar behavior was found in our previous work on {\Crtwo},\cite{Purwanto2015_Cr2}
where a multi-determinant truncated CASSCF trial wave function was also required.
In the more strongly correlated {\Crtwo}
molecule, however, AFQMC/UHF overestimated the binding by $\sim 0.9$\,eV near the equilibrium bond length, for a TZ
basis.\cite{Purwanto2015_Cr2}
For comparison, the CASPT2 results\cite{Borin2008} in the same basis set are also shown
in Fig.~\ref{fig:Mo2-PEC}.
We see that both the ground and excited states appear to be overbound, by $\sim 0.3$\,eV.
In the ground state of {\Crtwo}, CASPT2 results showed the same trend, resulting in
overbinding by $\sim 0.8$\,eV.\cite{Purwanto2015_Cr2,Kurashige2011}

\begin{figure}[!htbp]
\includegraphics[scale=0.33]{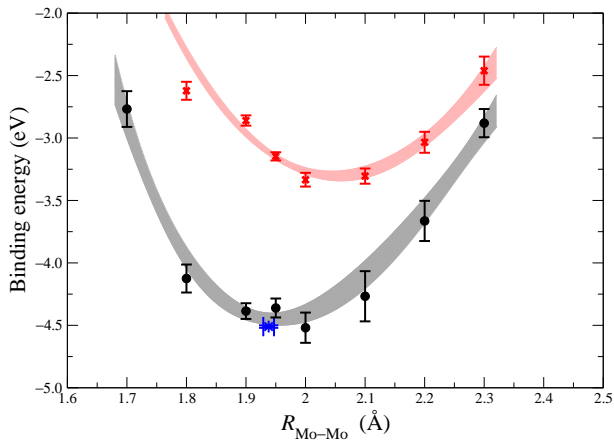}
\caption{\label{fig:Mo2-final-cbs}
CBS-extrapolated AFQMC/CASSCF PECs. Symbols, colors and shading are as in
Fig.~\ref{fig:Mo2-PEC}.
The ground state experimental binding energy and bond length are given by
the blue starred symbol, with the error bars representing the experimental uncertainties.
The zero-point energy has been removed from the experimental binding energy.
}
\end{figure}

The CBS-extrapolated AFQMC/CASSCF PECs are shown in \Fig{fig:Mo2-final-cbs}.
As in our previous work,\cite{Purwanto2015_Cr2}
the QZ$\to$CBS correction was obtained from AFQMC/UHF calculations using
the ANO-TZ (7s6p4d2f1g) and ANO-QZ basis sets.
We use a two-part scheme to extrapolate the many-body energies to the
CBS limit:\cite{Purwanto2011}
the exponential ansatz \cite{Halkier1999}
for the HF energies (with exponent $c=1.63$)
and the inverse cubic form\cite{Helgaker1997} for the correlation energies.
The QZ$\to$CBS correction increases the magnitude of the AFQMC binding energy
by about $0.3$\,eV and $0.2$\,eV at the shortest and longest bond
distances ($R_\textrm{Mo--Mo} = 1.7$ and $2.2$\,\AA, respectively).
In this geometry range, the correction is well approximated by a linear function of
$R_\textrm{Mo--Mo}$.
This correction was applied to ANO-QZ AFQMC/CASSCF PECs to obtain the CBS limit.
Test calculations showed that the excited state CBS correction
was within error bars of the ground state CBS value, so
we used the ground state correction for both.

Spectroscopic constants corresponding to Figs.~\ref{fig:Mo2-PEC} and  \Fig{fig:Mo2-final-cbs}
are given in Tables~\ref{tbl:Mo2-consts-v2} and
and  \ref{tbl:Mo2-consts-triplet-v2} for the ground and excited states, respectively.
The tables also show results from experiment and from other high-level quantum chemistry many-body calculations.
Our  coupled cluster singles and doubles with perturbative triples [CCSD(T)]
results were extrapolated to the CBS limit following the same procedure as described above, using CCSD(T) calculations for all basis sets;
the CBS correction obtained this way was slightly smaller than,
but consistent with, AFQMC/UHF CBS correction.
For multi-reference perturbative calculations, the CBS extrapolation  is less straightforward,
since their correlation energies do not fit the inverse-cubic ansatz well,
as discussed further below. Consequently, we made no attempt to apply the
same CBS corrections to the
perturbative results in Tables~\ref{tbl:Mo2-consts-v2} and
and  \ref{tbl:Mo2-consts-triplet-v2}; the values are listed in a separate column
and  correspond to the specified basis set.

As discussed in connection with Fig.~\ref{fig:Mo2-PEC}, AFQMC/CASSCF is essentially exact near equilibrium.
Thus, the AFQMC/CASSCF results in Tables~\ref{tbl:Mo2-consts-v2}
and  \ref{tbl:Mo2-consts-triplet-v2} provide a benchmark for assessing the other quantum chemistry methods. At the CBS limit AFQMC/CASSCF is seen to be in excellent agreement with experiment.
In contrast, other quantum chemistry results show considerable variance,
especially for the molecular dissociation energy $D_e$. This is discussed further in the next section.

\begin{table}[tbp]
\caption{\label{tbl:Mo2-consts-v2}
Ground state ($X ^1\Sigma_g^+$) spectroscopic properties of {\Motwo}
computed using phaseless AFQMC and other quantum chemistry methods.
$D_e$ is the molecular dissociation energy, in units of eV
(where the zero-point energy $\sim 0.03$ eV has been removed from the experimental value);
$R_0$ is the equilibrium bond length (in \AA);
and $\omega_e$ is the harmonic vibrational frequency (in cm$^{-1}$).
Unless otherwise indicated, the ANO-QZ basis
(see text) is used.
CBS extrapolation of $D_e$ is shown also, when applicable.
}
{%
\newcommand\Xj{\footnote{PT2 and PT3 refer to the second- and third-order perturbation theory, respectively.
NEV stands for the $N$-electron valence variant of the perturbation theory.}}
\newcommand\Xa{\footnote{\label{fn:t1:Xa}\Ref{Balasubramanian2002}.
Calculations use an ECP with 5s5p4d1f basis (see cited article for detail).}}
\newcommand\Xg{\footnote{\label{fn:t1:Xg}\Ref{Borin2008}.}}
\newcommand\Xh{\footnote{\label{fn:t1:Xh}\Ref{Angeli2007}.}}
\newcommand\Xb{\footnote{\label{fn:t1:Xb}CASPT2(12e,12o) method.}}
\newcommand\Xc{\footnote{\label{fn:t1:Xc}Strongly-contracted NEVPT2(12e,12o) method.}}
\newcommand\Xd{\footnote{\label{fn:t1:Xd}Basis set: ANO-QZ basis without the h functions (8s7p5d3f2g)}}
\newcommand\Xe{\footnote{\label{fn:t1:Xe}Basis set: full ANO basis (10s9p9d6f4g2h)}}
\newcommand\Xf{\footnote{\label{fn:t1:Xf}Strongly-contracted NEVPT3(12e,12o) method.}}
\newcommand\Ya{\footnote{\Ref{Simard1998}}}
\newcommand\Yb{\footnote{\Ref{Hopkins1983}}}
\newcommand\Yc{\footnote{\Ref{Efremov1978}}}
\newcommand\Noval{\;\;--}
\newcommand\fnref[1]{\textsuperscript{\ref{fn:t1:#1}}}
\begin{ruledtabular}
\begin{tabular}{lcllll}
Method              & & $D_e$    & $D_e$ (CBS)     & $R_0$        & $\omega_e$   \\
\hline
\multicolumn{4}{l}{Multireference perturbation theory} \\
\quad PT2--NEV (larger basis)\Xh\Xc\Xe & \COMMENTED{SC-NEVPT2(12e,12o), ANO-RCC full (10s9p9d6f4g2h)}
                      & $4.8845$ & \Noval          & $1.9187$     & $507.64$                \\
\quad PT2--NEV\fnref{Xh}\fnref{Xc}\Xd & \COMMENTED{SC-NEVPT2(12e,12o), ANO-RCC QZ$'$ (8s7p5d3f2g)}
                      & $5.055$  & \Noval          & $1.9198$     & $506.09$                \\
\quad PT3--NEV\fnref{Xh}\fnref{Xd}\Xf & \COMMENTED{SC-NEVPT3(12e,12o), ANO-RCC QZ$'$ (8s7p5d3f2g)}
                      & $3.9868$ & \Noval          & $1.9500$     & $461.54$                \\
\vspace{0.9ex}
\quad PT2\Xg\Xb  & \COMMENTED{Roos' CASPT2(12e,12o) ANO-RCC QZ}
                      & $4.41$   & \Noval          & $1.950$      & $459$                   \\
MRSDCI+$Q$\Xa       & & $3.9$    & \Noval          & $1.993$      & $447.5$                 \\
\vspace{0.9ex}
CCSD(T)             & & $3.85$   & $ 4.06 $        & $1.913$      & $549$                   \\
AFQMC/UHF           & & $4.45(1)$   & $ 4.66(1) $  & $1.955(4)$   & $428(5)$                \\
\vspace{0.9ex}
AFQMC/CASSCF        & & $4.20(5)$   & $ 4.46(5) $  & $1.95(2)$    & $467(24)$               \\
Experiment          & &            &  $4.51(1)$\Ya & $1.940(9)$\Yb & $477.1$\Yc             \\
\end{tabular}
\end{ruledtabular}
}
\end{table}
\begin{table}[tbp]
\caption{\label{tbl:Mo2-consts-triplet-v2}
Excited state ($a ^3\Sigma_u^+$) spectroscopic properties of {\Motwo}
computed using phaseless AFQMC and other quantum chemistry methods.
$T_e$ is the excitation energy from the ground state (in units of eV);
$R_0$ is the bond length at the PEC minimum (in \AA);
and $\omega_e$ is the harmonic vibrational frequency (in cm$^{-1}$).
Unless otherwise indicated, the ANO-QZ basis
(see text) is used.
CBS extrapolation of $T_e$ is shown also, when applicable.
}
{%
\newcommand\Xa{\footnote{\label{fn:t2:Xa}\Ref{Balasubramanian2002}.
Using an ECP with 5s5p4d1f basis (see cited article for detail).}}
\newcommand\Xg{\footnote{\label{fn:t2:Xg}\Ref{Borin2008}.}}
\newcommand\Xh{\footnote{\label{fn:t2:Xh}\Ref{Angeli2007}.}}
\newcommand\Xb{\footnote{\label{fn:t2:Xb}CASPT2(12e,12o) method.}}
\newcommand\Xc{\footnote{\label{fn:t2:Xc}Strongly-contracted NEVPT2(12e,12o) method.}}
\newcommand\Xd{\footnote{\label{fn:t2:Xd}Basis set: ANO-QZ basis without the h functions (8s7p5d3f2g)}}
\newcommand\Xe{\footnote{\label{fn:t2:Xe}Basis set: full ANO basis (10s9p9d6f4g2h)}}
\newcommand\Xf{\footnote{\label{fn:t2:Xf}Strongly-contracted NEVPT3(12e,12o) method.}}
\newcommand\Ya{\footnote{\label{fn:t2:Ya}\Ref{Krauss2001}}}
\newcommand\Yb{\footnote{Ref{XXX}}}
\newcommand\Yc{\footnote{Ref{XXX}}}
\newcommand\Noval{\;\;--}
\newcommand\fnref[1]{\textsuperscript{\ref{fn:t2:#1}}}
\begin{ruledtabular}
\begin{tabular}{lcllll}
Method              & & $T_e$    & $T_e$ (CBS)     & $R_0$        & $\omega_e$   \\
\hline
PT2\Xg\Xb  &
                      & $1.105$  & \Noval          & $2.063$      & $393$                   \\
\vspace{0.9ex}
\vspace{0.9ex}%
AFQMC/CASSCF        & & $1.15(6)$   & $1.15(6)$    & $2.05(1)$    &$399(20)$               \\
Experiment          & &          &  $0.9947$\Ya    &  \Noval      & $393.7$\fnref{Ya}      \\
\end{tabular}
\end{ruledtabular}
}
\end{table}

\section{Discussion}
\label{sec:Discussion}

In this section, we analyze in more detail
both the comparison of AFQMC with other methods
and the comparison of the correlation effects in {\Motwo} with {\Crtwo}.
Correlation effects are greatly reduced in {\Motwo} compared to {\Crtwo}.
While CASSCF result is not bound for {\Crtwo}
for an active space as large as (12e,28o),
in {\Motwo} CASSCF(12e,12o) already recovers $30\%$ of the binding energy.
For systems with strong static correlation,
multireference perturbation theory has often been the method of choice.
As shown in Table~\ref{tbl:Mo2-consts-v2}, however,
the results depend rather sensitively on the perturbative implementation.
Given that {\Motwo} is considerably more benign than {\Crtwo},
it is perhaps somewhat surprising that it turns out to be  rather challenging to the best quantum
chemistry methods.

\subsection{Comparison with other theoretical results}

The CASPT2 calculation of Borin \emph{et al.}\cite{Borin2008},  using the ANO-QZ basis and a CASSCF(12e,12o) active space zero-order wave function,
overestimates $D_e$
by $\sim 0.2$\,eV.
In {\Crtwo}, CASPT2(12,12) shows much larger overbinding of \mbox{$\sim 1.0$\,eV}, based on direct comparisons with the exact
FP-AFQMC
\cite{Purwanto2015_Cr2}.
This trend has been attributed to the inadequacy of the active space chosen for the
zeroth-order wave function.\cite{Kurashige2011,Angeli2007}
In {\Crtwo}, an improved zeroth-order wave function was obtained for a larger active space of
12 electrons and 28 orbitals, using a self-consistent density matrix renormalization group
(DMRG) calculation.\cite{Kurashige2011}
The subsequent CASPT2 calculation \cite{Kurashige2011} reduces, but does not eliminate, the discrepancy with exact FP-AFQMC, resulting
in underbinding by \mbox{$\sim 0.4$\,eV}.\cite{Purwanto2015_Cr2}
Multireference perturbative methods are also sensitive to
the perturbative implementation.
Results from an alternative perturbation treatment, using $n$-electron valence
perturbation theory (NEVPT)\cite{Angeli2002} are also shown in Table~\ref{tbl:Mo2-consts-v2}.
The second-order PT2-NEV results show larger overbinding,
while third-order PT3-NEV is underbound by \mbox{$\sim 0.2$\,eV}.
The $D_e$ of PT2-NEV is reduced by $0.2$\,eV upon increasing the ANO basis size.
This is opposite to AFQMC and CCSD(T), which show $D_e$ increasing with basis size.
The basis convergence of the perturbative calculations does not follow
the empirical $x^{-3}$ behavior, where $x$ is the
(correlation consistent) basis cardinal number.
(Hence no CBS extrapolation is performed on the results, as mentioned earlier.)
In view of these considerations, the
agreement of the CASPT2 {\Motwo} ANO-QZ calculation
with the experimental binding energy is likely somewhat fortuitous.

Table~\ref{tbl:Mo2-consts-v2} also presents results from multireference
singles and doubles configuration interaction (MRSDCI) calculations  \cite{Balasubramanian2002},
with $Q$ correction applied, and from CCSD(T).
The MRSDCI+$Q$ used an
ECP and a smaller basis.
Single-reference CCSD(T), with restricted HF reference wave function,
performs much better in {\Motwo} than in {\Crtwo}.
While CCSD(T) predicts that
the {\Crtwo} molecule is unbound,
for {\Motwo} it yields respectable agreement with experiment.
Nevertheless, the multireference character of the {\Motwo} ground state
is sufficiently strong that CCSD(T) still underestimates $D_e$ by $0.45$\,eV,
as shown in Table~\ref{tbl:Mo2-consts-v2}.

For the  {\Motwo} ground state, all of the standard methods have difficulty
obtaining an accurate dissociation energy.
For the triplet excited state (Table~\ref{tbl:Mo2-consts-triplet-v2}),
we find that the AFQMC and CASPT2 results agree very well with each other and with experiment.

\subsection{Comparing {\Crtwo} and {\Motwo}}

\begin{figure}[tbp]
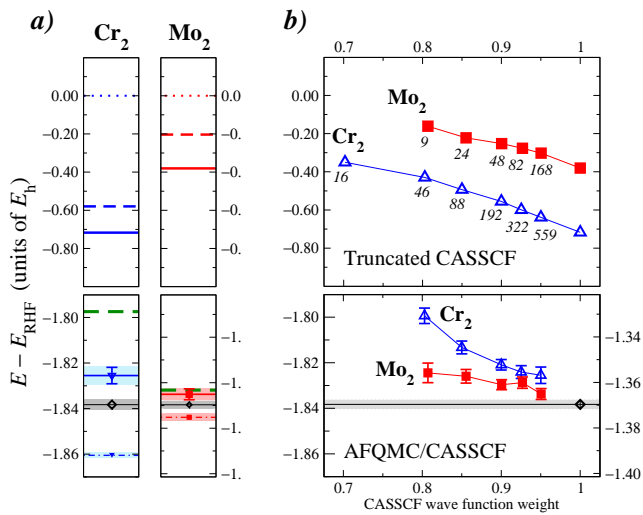

\includegraphics[scale=0.34,valign=t]{\FigureFile{Mo2-vs-Cr2-corr-v2}}%
\quad%
\includegraphics[scale=0.34,valign=t]{\FigureFile{Mo2-vs-Cr2-corr-v2p2}}

\caption{\label{fig:Cr2-Mo2-corr}
The correlation energy in the {\Crtwo} and {\Motwo} molecules.
All energies are reported relative to the restricted HF (RHF) energy.
Note that the energy offset in the lower panels of (a) and (b) aligns the (exact) FP-AFQMC correlation energies of {\Crtwo} and {\Motwo}.
\mbox{a) Upper panels:} RHF (dotted zero baseline), UHF (dashed line), and CASSCF (solid line) energies.
Lower panels: CCSD(T) (dashed line), AFQMC/CASSF at 95\% weight cutoff (triangle and square), FP-AFQMC (diamond), AFQMC/UHF (circle); the AFQMC statistical uncertainties are indicated by the shading.
b)~Correlation energy as a function of
the retained weight in the multi-determinant CASSCF wave function.
Solid squares and open triangles denote the {\Motwo} and {\Crtwo}
results, respectively.
Upper panel: variational energy 
of the truncated CASSCF wave function; numbers adjacent to the symbols
give the number of determinants in the truncated wave function.
\mbox{Lower panel:} the corresponding AFQMC/CASSCF correlation energies; statistical uncertainties indicated by the error bars.
The exact FP-AFQMC energy from (a) is also shown.
See the text for additional details.
}
\end{figure}

In this section, we quantitatively compare the effect of electron-electron correlation in
{\Crtwo} and {\Motwo}. We use the exact FP-AFQMC to benchmark the relative effects between the two molecules.
Figure~\ref{fig:Cr2-Mo2-corr} compares the magnitude of electron-electron correlation effects
in the {\Crtwo} and {\Motwo} molecules.
Results for {\Crtwo} were obtained using the cc-pwCVTZ-DK basis
at the experimental
bond length \mbox{$R = 1.6788\,$\AA{}}
(see \Ref{Purwanto2015_Cr2});
the {\Crtwo} RHF energy is $-2098.533662~\Eh$.
Results for {\Motwo} were obtained using the ANO-QZ basis
near the experimental
bond length \mbox{$R = 1.9\,$\AA};
the {\Motwo} RHF energy is $-8091.069911~\Eh$.
Although the correlation energy is $\sim 0.5\,\Eh$ larger in {\Crtwo} than in {\Motwo},
the  {\Crtwo} UHF wave function recovers a larger fraction of the correlation energy, $32\%$, versus $15\%$ in {\Motwo}.
The CASSCF wave function shows a similar but less pronounced trend, recovering  $39\%$ and $28\%$ in {\Crtwo} and {\Motwo}, respectively.

The stronger correlation effects in {\Crtwo}, however,
are evident in the  top panel of Fig.~\ref{fig:Cr2-Mo2-corr}b:
achieving $95\%$ of the CASSCF total wave function weight requires retaining
$559$ determinants in {\Crtwo} but  only $168$ in {\Motwo}.
This is also evident in the larger CCSD(T) discrepancy in {\Crtwo} than in {\Motwo},
$\sim 40\,\mEh$ and $\sim 7\,\mEh$, respectively.
The dependence of the AFQMC/CASSCF energies
on the quality of the trial wave function is shown in the bottom panel of Fig.~\ref{fig:Cr2-Mo2-corr}b.
The dependence is significantly stronger in {\Crtwo}, where at $95\%$
cutoff, the total energy is still $\sim 13\,\mEh$ higher than the exact
value (this error is $\sim 5\,\mEh$ for {\Motwo}).
At the variational level, although the truncated wave function recovers more correlation
energy in {\Crtwo} than in {\Motwo} for the same weight cut, its performance is worse in AFQMC/CASSCF.
We attribute this to the larger
dynamic correlation energy
that must be recovered in {\Crtwo}.
Cancellation of errors between the molecule and atom AFQMC energies
leads to better agreement in binding energy, however. For the best truncated CASSCF
wave function, the error in the binding energy
is $\sim 5\,\mEh$ ($\sim 0.14$\,eV) for {\Crtwo},
and virtually exact for {\Motwo}.

Basis set errors will modify the correlation energy recovered by the different methods. The mean-field HF energies are quite
well converged for the basis sets used here. For the many-body calculations,
we estimate (using AFQMC/UHF) the CBS shifts to be $\sim -100\,\mEh$ and $\sim -70\,\mEh$ for {\Crtwo} and {\Motwo},
respectively. For the purpose of the above comparisons, however, the relative error between the various approximate and exact methods
should not change significantly.

\section{Summary}
\label{sec:Summary}

We have presented an accurate calculation of
the {\Motwo} ground state ($X\,^1\Sigma_g^+$) and
first triplet excited state ($a\,^3\Sigma_u^+$).
We use the phaseless AFQMC method with the truncated CASSCF trial wave function
(AFQMC/CASSCF).
Calculations were done using high-quality, realistic basis sets,
and extrapolation to the CBS limit is performed.
The resulting PECs and spectroscopic constants are
in excellent agreement with experiment.
Comparisons are made with other high-level quantum chemistry methods.
We have also quantified the extent of strong electron correlations in both
{\Crtwo} and {\Motwo} molecules.
Molybdenum is important in a variety systems which can potentially exhibit exotic properties from strong correlation and topological effects.
Our results can serve as a benchmark as theoretical and computational methods are developed and employed to treat such systems reliably.

\begin{acknowledgments}

This work was supported by
DOE (DE-FG02-09ER16046),
NSF (DMR-1409510),
and
ONR (N000141211042).
We acknowledge a DOE CMCSN grant (DE-FG02-11ER16257)
for facilitating stimulating interactions.
An award of computer time was provided by
the Innovative and Novel Computational Impact on Theory and Experiment
(INCITE) program,
using resources of the Oak Ridge Leadership Computing Facility (Titan)
at the Oak Ridge National Laboratory,
which is supported by the Office of Science of the U.S. Department of Energy
under Contract No. DE-AC05-00OR22725.
We also acknowledge the computational support provided by
the William and Mary High Performance Computing facility.
The authors would like to thank
Valera Veryazov for clarifying the ANO basis, and
Fengjie Ma and Hao Shi for
many useful discussions.

\end{acknowledgments}

\bibliography{AFQMC-bib-entries,Mo2}

\end{document}